\documentclass[twocolumn]{svjour3}          
\usepackage{graphicx}
\usepackage{bm}
\usepackage{amsmath} 
\def\cm{cm$^{-1}$}
 \journalname{Theoretical Chemistry Accounts}
\begin{document}
\title{Properties of Hydrogen Bonds in the Protic Ionic Liquid  Ethylammonium Nitrate}
\subtitle{DFT versus DFTB Molecular Dynamics}
\titlerunning{H-bond Dynamics of EAN}        
\author{Tobias Zentel        	 \and
        Oliver K\"uhn 
}


\institute{Tobias Zentel \at
Institut f\"ur Physik \\ 
Universit\"at Rostock \\
Albert-Einstein-Str. 23-24 \\
18059 Rostock, Germany  \\
\email{tobias.zentel@uni-rostock.de}           
           \and
Oliver K\"uhn \at
Institut f\"ur Physik \\ 
Universit\"at Rostock \\
Albert-Einstein-Str. 23-24 \\
18059 Rostock, Germany  \\
\email{oliver.kuehn@uni-rostock.de}           
}

\date{Received: date / Accepted: date}
\maketitle
\begin{abstract}
  Comparative molecular dynamics  simulations of a hexamer cluster of the  protic ionic liquid ethylammonium nitrate are performed using density functional theory (DFT) and density functional-based tight binding (DFTB) methods. The focus is on assessing the performance of the DFTB approach to describe the dynamics and infrared spectroscopic signatures of hydrogen bonding between the ions. Average geometries and geometric correlations are found to be rather similar. The same holds true for the far-infrared spectral region. Differences are more pronounced  for the NH- and CH-stretching band, where DFTB predicts a broader intensity distribution. DFTB completely fails to describe the fingerprint range shaped by nitrate anion vibrations. Finally, charge fluctuations within the H-bonds are characterized yielding moderate dependencies on geometry. On the basis of these results, DFTB is recommend for the simulation of H-bond properties of this type of ionic liquids.

\end{abstract}
\section{Introduction}
\label{intro}
Ionic liquids (ILs) are promissing candidates for novel applications, including the use as solvents and CO$_2$ absorbers~\cite{Firaha15_7805,Kerle09_12727,Welton99_2071}. Their special physico-chemical pro\-perties are determined by the  nanoscopic structure and dynamics, which are governed mainly by strong Coulomb forces between the ions. For example, the ion pair lifetime can be linked directly to transport properties as shown by Zhang and Maginn~\cite{Zhang15_700}. There also exist directional hydrogen bond (HB) interactions between ions of opposite charge within a wide range of HB strengths~\cite{hunt15_1257}. Although being much weaker compared to Coulomb interactions, they impact macroscopic properties decisively. A prominent example is the reduction of the viscosity by the presence of HBs as shown in Ref.~\cite{Fumino08_8731}. 

Experimentally, HB interactions are accessible by far infrared (FIR) spectroscopy, as was discussed in detail by Fumino \textit{et al.}~\cite{Fumino12_6236}, as an extra blue-shifted band if compared with the effect of dispersion interaction. Furthermore, in the mid-infrared (MIR) region the H-bonded NH- or CH-stretching vibrations are red-shifted with respect to their non-H-bonded counterparts~\cite{Roth12_105026}. The signatures of H-bonding and Fermi resonance interactions can be investigated in detail using nonlinear spectroscopic techniques, which are capable to unravel different phase and energy relaxation rates~\cite{nibbering07_619,Zhuang09_3750,Chatzipapadopoulos15_2519}. 

Numerical simulations can be used to support the various spectroscopic evidences for the existence of HB interactions. Thereby, one can distinguish between cluster and bulk liquid setups.   Although these systems are rather different at first glance, gas phase spectroscopy of ILs in dependence on cluster size has revealed  local motifs  that give rise to   bulk like spectra \cite{Johnson13_224305}. Structural similarities have also been found in simulations and were assigned to the dominance of  strong short-range Coulomb interactions~\cite{Bodo13_144309}.  Cluster models are particularly well-suited for a normal mode analysis, corresponding to a minimum structure at zero Kelvin. The resulting frequencies are commonly scaled by a method-dependent factor    and the spectra are artificially broadened   to fit the experimental line shapes. Still, the pronounced anharmonicity of HB dynamics might require to calculate explicit potential energy surfaces~\cite{giese06_211}.  In addition, there exist trajectory-based  normal modes approaches that   can account for temperature effects~\cite{mathias11_2028}, which also have been applied to ILs~\cite{Thomas14_024510}. However, the standard approach especially for bulk simulations makes use of linear response theory, \textit{i.e.} the IR spectrum is calculated by Fourier transformation of the dipole autocorrelation function~\cite{may11}. The latter is obtained by sampling MD trajectories from a canonical ensemble~\cite{ivanov13_10270}. Of course,  the quality of the  spectra depends strongly on the underlying potential energy surface driving the dynamics.

Classical MD simulations use many body interactions parametrized via force fields. There are many applications to structural and dynamical properties of ILs \cite{Maginn09_373101,DelPopolo04_1744}, discovering, for example, mesoscopic segregation behavior~\cite{Bernardes14_6885,Shimizu14_567}. Concerning IR spectroscopy and in particular signatures of HB dynamics the use of force fields is problematic, since the parametrization of the latter is often targeted to thermodynamic quantities, see, e.g., Ref.~\cite{zentel16_234504}. Here, DFT-based \textit{ab initio} molecular dynamics (AIMD) provides the proper frame for more accurate simulations~\cite{Marx10_,Bodo13_144309,kirchner15_202}. For instance, the structure and HB dynamics of methylammonium nitrate was investigated in detail by Zahn \textit{et al.} using  Car-Parinello MD \cite{Zahn10_124506}. Employing radial, angle and spatial distribution functions, preferred ion orientations and HB properties were investigated and the behavior of ion caging was observed. The latter has been discussed as a general feature of IL dynamics~\cite{Schroder11_024502}. For small system sizes it is in principle possible to obtain AIMD trajectories long enough to calculate the power spectra \cite{Wendler12_1570} through velocity autocorrelation functions or, in combination with a charge localization scheme, IR spectra via dipole autocorrelation functions \cite{Thomas15_3207}. 

Density functional-based tight-binding (DFTB) is a computationally less demanding alternative, thus allowing for larger system sizes \cite{Elstner98_7260,Yang07_10861}.  DFTB doesn't require empirical input and provides self-consistent Mulliken charges, which account for polarization effects. It has been successfully applied to a large class of problems, including biological systems \cite{Elstner06_316}, as well as to study structural properties of ILs~\cite{zentel16_234504,Addicoat14_4633}. 

In this contribution we address the performance of DFTB with respect to its description of HB dynamics and IR spectra.  This is done for the test system  ethylammonium nitrate (EAN), [(C$_2$H$_5$)NH$_3]^+$[NO$_3]^-$, which is a prototypical protic ionic liquid (PIL). The reliability of the results will be judged against data obtained by DFT/BLYP AIMD simulations. To facilitate the AIMD calculations the system is restricted to a hexamer cluster of six ion pairs. Previously, DFT and force field calculations have been compared for EAN clusters of various sizes by  Bodo \textit{et al.}~\cite{Bodo12_13878,Bodo13_144309}. These authors put a special emphasis on HB geometries of \emph{energy optimized structures}. In DFT structures they found asymmetric HB networks  with  only two out of three possible HB formed. In contrast force field simulations yielded more symmetric structures and saturated HBs. Furthermore, they compared a harmonic normal mode spectrum  against Raman measurements and concluded on signals from NH-stretching vibrations involved in HBs of different strength. The force field itself had been parametrized by Song \textit{et al.}  by combining various parameter sets with additional \textit{ab initio} data~\cite{Song12_2801}. Radial distribution functions from MD simulations were found in reasonable agreement with high-energy X-ray diffraction data. Interestingly,  OH-distances and HB  angles in force field MD simulations are substantially larger than the DFT ones from energy optimized structures~\cite{Bodo12_13878} and from later AIM simulations~\cite{Bodo13_144309}.

DFTB has been compared to force field simulations for the related PIL triethylammonium nitrate in our previous work~\cite{zentel16_234504}. Preceding bulk simulations, geometric HB correlations for a hexamer cluster have been compared for DFTB and DFT trajectories. This gave support for the reliability of the DFTB method, which then was used to quantify geometric correlations and to obtain IR spectra for the liquid phase. In particular for the FIR range, where experimental data are available, DFTB yielded excellent results whereas the force field failed to reproduce the experiments. 

Alkylammonium cations provide the means to tune the network of HBs formed between cations and anions via the number of alkyl chains~\cite{fumino09_3184,zentel16_}. Whereas triethylammonium nitrate is capable of forming a single HB only, the present system, EAN, in principle facilitates the formation of three HBs. In the following the HB geometries and IR signatures of EAN will be investigated using cluster MD simulations. Thereby, DFTB and DFT trajectory data will be compared, which will lend support for DFTB as an efficient yet accurate method for simulation of the HB network in PILs such as EAN. The paper is organized as follows: In the next section a brief overview on the used theoretical methods and computational setups is given. Results are presented in  Section \ref{sec:results}, focussing on geometries, IR spectra, and charge distributions. Finally, a conclusion is provided in Section \ref{sec:conclusions}.

\section{Theoretical Methods}
\subsection{Molecular Dynamics}
\begin{figure}
\begin{center}
\includegraphics[width=0.5 \textwidth]{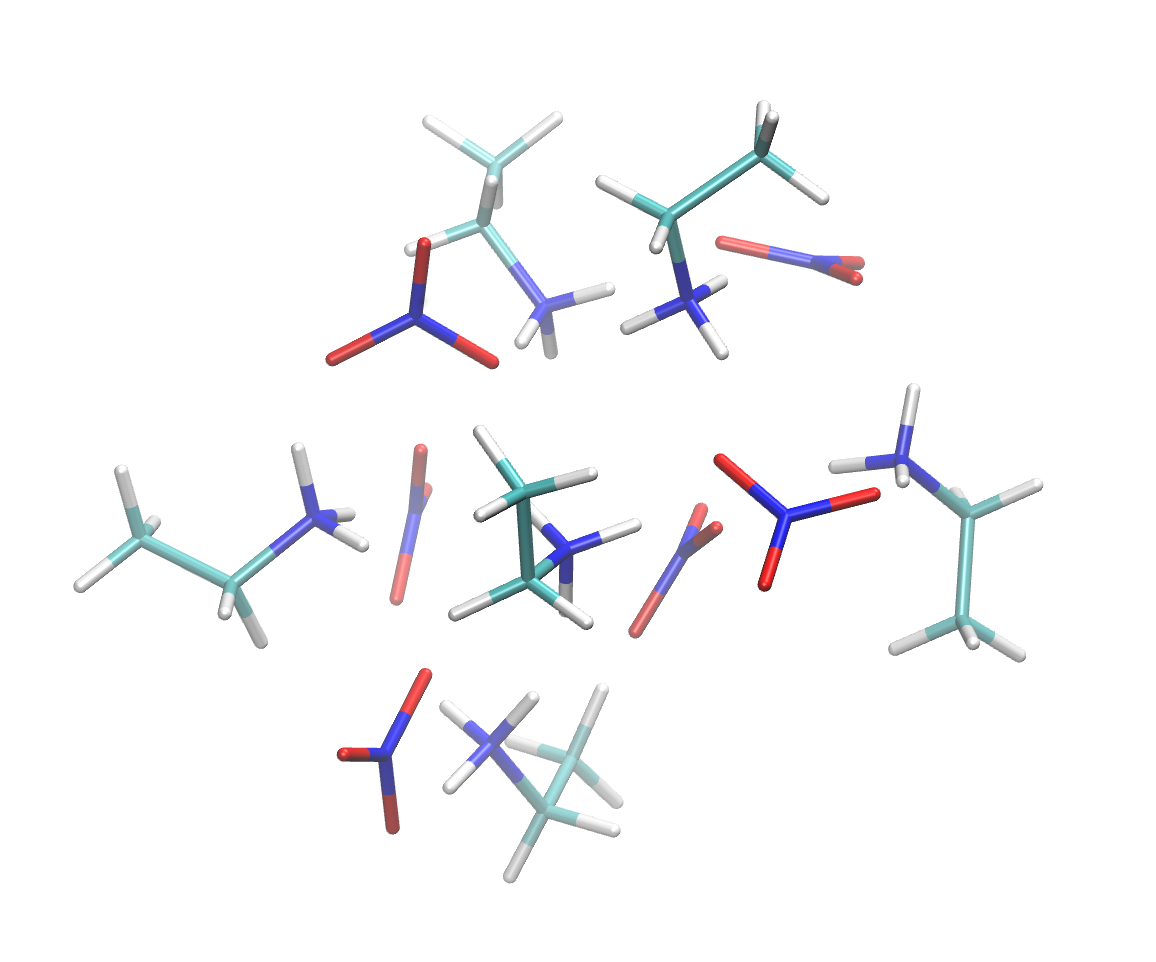}
\caption{ Snapshot along a NVE DFTB trajectory of the hexamer cluster consisting of six ethylammonium nitrate (EAN), [(C$_2$H$_5$)NH$_3]^+$[NO$_3]^-$, ion pairs. The figure has been generated using VMD~\cite{Humphrey96_33}.}
\label{fig:Geocluster}
\end{center}
\end{figure}
A cluster consisting of 90 atoms (six ion pairs) with similar initial structure as the hexamer cluster studied by Bodo \textit{et al.}~\cite{Bodo12_13878}, is used for the following DFTB and AIMD simulations.  In a first step the cluster has been geometry-optimized using DFT with the BLYP exchange correlation functional and the 6-31+G* basis set. Further, Grimme's D3 dispersion correction has been employed \cite{Grimme10_154104}.  Starting from the minimum structure, Langevin dynamics was performed with the target temperature being 300~K and a Langevin damping time of 1~ps. The stochastic perturbations introduced by the Langevin thermostat are small enough to yield spectra equivalent to the ones from the standard approach (microcanonical trajectories  sampled from an canonical ensemble) \cite{Gottwald12_}. The  total NVT trajectory length has been 60~ps and the  time step  0.4~fs. For calculating the IR spectrum, it is split into trajectories of 10~ps length.  The DFT AIMD simulations, in the following called BLYP,  were performed with Terachem, version 1.5K~\cite{Ufimtsev09_2616}.

Starting from the same initial BLYP geometry, DFTB simulations were done using third-order density expansion~\cite{Gaus11_931} and the 3-ob Slater-Koster parameter set~\cite{Gaus13_338}. The DFTB simulations are carried out using the dftb+ software, Version 1.2~\cite{Aradi07_5678}. First, a canonical ensemble was simulated for  5~ps employng the Nos\'e-Hoover thermostat with target temperature 300~K. This was followed by an energy optimization. Subsequently, the optimized structure is used to initiate  a canonical ensemble run for 100~ps. From this trajectory starting geometries for seven microcanonical (NVE) trajectories were sampled randomly. The microcanonical trajectories had a length of 10~ps and a time step of 0.4~fs.

DFT and DFTB trajectories were analyzed with respect to the presence of HBs. Here, a HB was defined between the nitrogen-bound hydrogen and its closest oxygen atom, without a distance cutoff. This way each of the six cations were assigned three HB partner oxygens from the anions. The meaningfulness of this choice is analyzed below.    

\subsection{IR Spectra} 
 The IR spectrum, $I(\omega)$, is obtained by Fourier transformation of  the dipole moment trajectory, \vec{\mu}(t), using a Kaiser window function, $\kappa(t)$, with Kaiser parameter 10, \textit{i.e.}
  \begin{equation}
I(\omega) = \omega^2 \langle | \int\limits_{0}^{T} \exp(-i \omega t)   \vec{\mu}(t) \kappa(t)   \mathrm{d}t \ |^2 \rangle \, .
\label{eq:IRsignal}
\end{equation}
Here, the average is taken with respect to the canonical ensemble. The window function ensures that the dipole fluctuations are damped to zero at the integration boundaries. The Kaiser parameter controls the time scale of the decay and is chosen large enough to ensure that the spectra are not artificially broadened. All spectra plotted are the moving averages of 21 steps to smooth the curves. In BLYP the total cluster dipole is calculated directly from the wave function, whereas in DFTB the Mulliken charges are used to calculate the total dipole trajectories of the cluster. Note that the  dipole in an ionic systems is not uniquely defined, as discussed in Ref.~\cite{Brehm12_5030}. In the context of IR spectroscopy the center of mass of the cluster is a plausible reference, as this point is fixed for all vibrational normal modes \cite{Thomas15_3207}. In addition to the total cluster dipole, the contribution of the $\mathrm{NH_3}$ group atoms to the total dipole are extracted from DFTB Mulliken charges.

For comparison vibrational normal modes are calculated for both methods. In case of DFT the Turbomole software package has been used~\cite{Ahlrichs89_165,turbomole6.5,Deglmann04_103}. 
\subsection{Population Analysis} 
As compared to standard force fields DFTB has the advantage of delivering atom-centered  point charges (Mulliken charges), which are not only determined self-consistently, but they are polarizable in the sense that they adopt to the actual environment of the atom~\cite{zentel16_234504}.

In general the partial charge of an atom $A$ with atomic number $Z_A$ is defined by
\begin{equation}
q^{A} = Z_A - \int \rho_A(r) \mathrm{d}r . 
\end{equation}
Depending on the method used the partial charges can vary because the atomic densities, $\rho_A(r)$, can be defined in different ways. A simple approach to obtain partial charges is the Mulliken population analysis. Here, the total electron density, $\rho(r)$, is written in terms of the density matrix elements, $P_{ab}$, and the atomic basis functions, which after integration over space gives the total number of electrons 
\begin{equation}
N = \int \rho(r) \mathrm{d}r = \sum_{ab } P_{ab} S_{ab} \ ,
\end{equation}
where $S_{ab}$ is overlap matrix element and the indices $a$ and $b$ refer to atomic basis functions. The Mulliken population $N_A$ of an atom $A$ is defined as 
\begin{equation}
N_A =\int \rho_{\mathrm{A}}(r) \mathrm{d}r =
 \sum_{a \in A} P_{a a}  +\frac{1}{2} \sum_{\substack{a  \in A \\ b \notin A}}  P_{a b}S_{ab} \, .
\end{equation} 
%
To gain insight into the  charge distribution within the HB, Mulliken analyses are performed for the respective N, O, and H atoms using a  set of methods, including DFTB, DFT, and coupled cluster theory (CCSD(T)); Turbomole is used for the CCSD(T) calculations \cite{Ahlrichs89_165,turbomole6.5,Haser89_104,Hatting00_5154}. 

\section{Results} 
\label{sec:results}
\subsection{Cluster Structure and HB Geometries}

\begin{figure}
\includegraphics[width= \columnwidth]{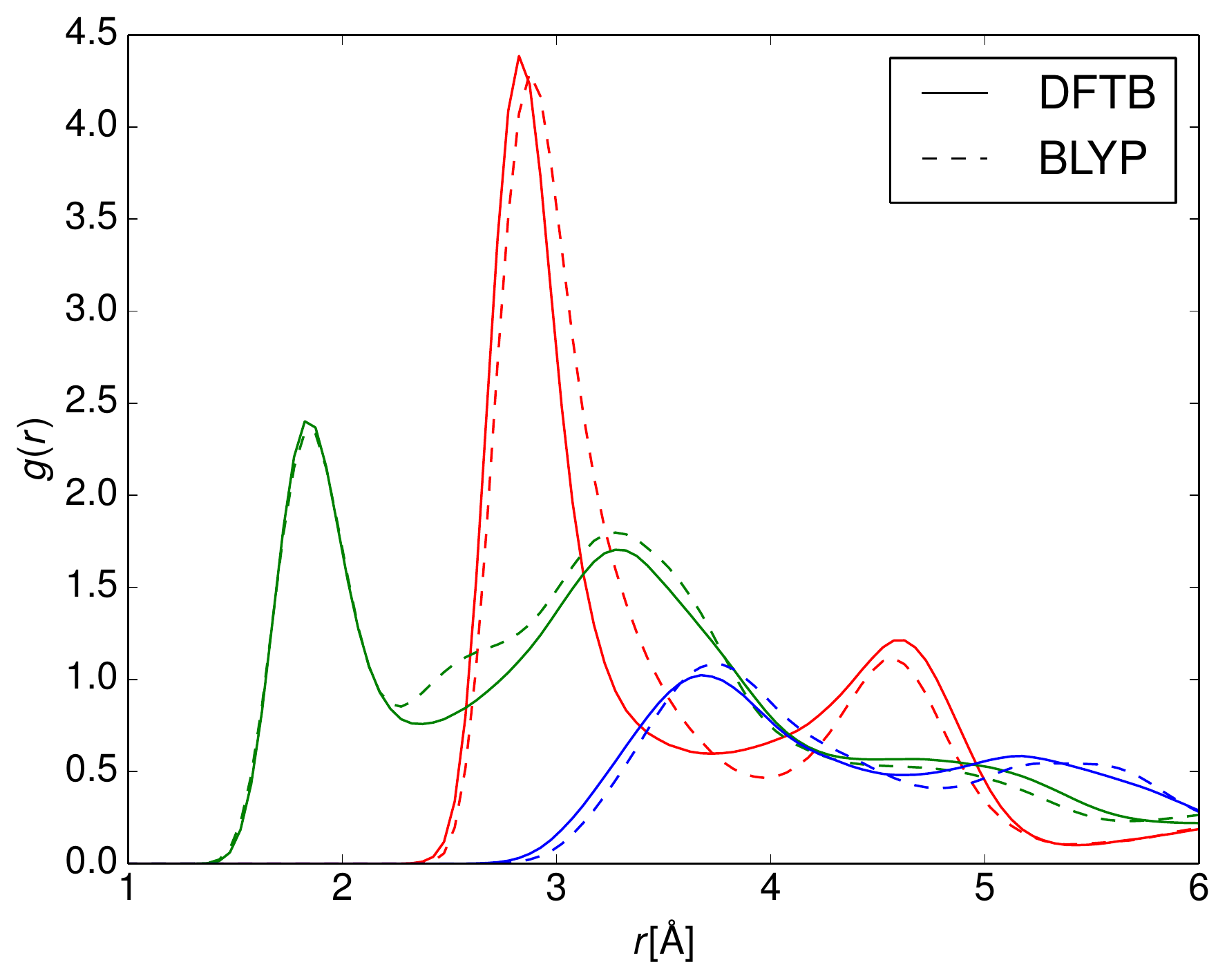}
\caption{Radial distribution functions $g_{\mathrm{ON}}(r)$ (red), $g_{\mathrm{OH}}(r)$ (green), and $g_{\mathrm{OC}}(r)$ (blue), with the solid and dashed lines corresponding to the  DFTB and BLYP case, respectively. Data have been extracted from the NVT trajectory every 2.5~fs.
\label{fig:rdf}}
\end{figure}

The radial pair distribution functions, $g_{\mathrm{ON}}(r)$, $g_{\mathrm{OH}}(r)$, and $g_{\mathrm{OC}}(r)$ have been calculated from the distances between the anionic O to the cationic N, the H bound to it, and the alkyl C atoms, respectively. DFTB and BLYP results  are compared in Fig.~\ref{fig:rdf}. The convergence with respect to the trajectory length was checked for the DFTB simulations. Note that due to the cluster structure, $g(r)$ approaches zero for larger distances. In general these pair correlation functions are rather similar for DFTB and DFT clusters. The largest difference between DFTB and DFT is visible for $g_{\mathrm{OH}}(r)$ around 2.5~$\mathrm{\AA}$, which is due to  slightly different orientations of the $\mathrm{NO}_3^-$ molecules with respect to the N-H bond, bringing a second O atom closer to the N-H bond (for values of first peak, see Tab.~\ref{tab:rdfpeaks}). 

HBs can be detected as distinct peaks at small distances between  two atoms involved in HB. Indeed, $g_{\mathrm{OH}}$ and  $g_{\mathrm{ON}}$  show maxima at short distances, but there is no such contribution in $g_{\mathrm{OC}}$. Thus HBs only develop via the cationic nitrogen atom to the anion, and not via the alkyl chains.

\begin{table}[h]
\begin{center}
\begin{tabular}{|l|c|c|c|c|}
\hline 
&  OH & ON & OC & $ \left< \alpha \right> \ [^{\circ}] $\\ 
\hline
\hline
BLYP   & 1.9 & 2.9 & 3.7 & 24.4\\ \hline
DFTB   & 1.9 & 2.8 & 3.7 & 29.7\\ 
\hline 
Ref.~\cite{Bodo13_144309} & 1.8 & 2.8 & -  & - \\ 
\hline
Ref.~\cite{Hayes11_3237}   & 2.4 & 3.0 &  - & -\\
\hline
Ref.~\cite{Addicoat14_4633} & - & 2.5 & 3.0 & -\\
\hline 
Ref.~\cite{Song12_2801}   & 2.5 & 3.2 & 3.5 & -\\  
\hline
\end{tabular}
\caption{Comparison of position of the first peak (in \AA) in the radial pair distribution function of the hexamer cluster as calculated with different methods. For comparison bulk neutron diffraction~\cite{Hayes11_3237}, DFTB~\cite{Addicoat14_4633}, force field MD~\cite{Song12_2801}, and Car-Parrinello (BP86) AIMD~\cite{Bodo13_144309}  results  are given as well. In the right-most column one finds the average HB angle.
} 
\label{tab:rdfpeaks}
\end{center}
\end{table}

Cluster results are available from the work of  Bodo \emph{ et al.}. In Ref.~\cite{Bodo12_13878} they reported distances obtained for an equilibrium (geometry optimized) hexamer cluster structure. Here, average OH bond length of 1.86~\AA{} and ON distances of 2.9~\AA{}(DFT/$\omega$B97X-D) have been found. Car-Parrinello AIMD simulations of the same hexamer~\cite{Bodo13_144309}  yielded rather similar values, i.e. 1.8~\AA{} and 2.8~\AA{} for the average OH and ON distance, respectively. These values actually agree very well with those of bulk phase simulations reported in the same reference. This has been  attributed to the similarity of the first solvation shell due to the prevalence of strong Coulomb interactions. There are other bulk phase investigations as well. For instance, an analysis of neutron diffraction data has been reported in Ref.~\cite{Hayes11_3237}; cf. Tab.~\ref{tab:rdfpeaks}. The ON distance compares rather well with the present findings. The OH distance is smaller by 0.5~\AA{}, both in DFTB and BLYP. Comparing with force field MD simulation data of Song \textit{et al.} \cite{Song12_2801} this could be a bulk effect, although such a conclusion would be at variance with the AIMD results of Ref.~\cite{Bodo13_144309}.  In this respect we note that the bulk DFTB results reported in Ref.~\cite{Addicoat14_4633} are markedly different from experimental and MD data. To conclude, the present DFTB setup yields average HB geometries, which are consistent with both BLYP and Car-Parrinello results.

%


As can be seen in Fig. \ref{fig:Geocluster} the cations orient themselves such that the ethyl chains are on the outer border of the cluster and the polar  head group is close to the anions and forms HB to oxygens from different anions. The HBs were identified  from the trajectory based on the shortest $r_{\mathrm{OH}}$  found, thus all nitrogen bound hydrogens are supplied with a H-bonding partner oxygen atom. To verify this procedure, all   HB pairs are analyzed as well by the heavy atom distance, $r_{\mathrm{NO}}$, combined with an angle criterion. Applying the  classifications of moderate HB strength of $r_{\mathrm{NO}} <3.2~\mathrm{\AA}$ and $\alpha <50^{\circ}$ from Ref.~\cite{hunt15_1257}, 15.6 \% and  10.1 \%  of the configurations are not HBs in DFTB and BLYP, respectively. Interestingly, in both cases in less than 1 \% of the samples distance and angle criteria are violated at the same time. Moreover, in 29\% (DFTB) or  25 \% (BLYP) of the cases one has twofold HBs, where one oxygen is acceptor of two H atoms. These twofold HBs do not show increased OH distances or HB angles, compared to the other HBs, suggesting that they are of similar strength. Thus we conclude that in the present cluster the NH groups are mostly saturated and form an asymmetric network where an O atom can partake in two HBs, leaving some O atoms without H-bonding partners. 

These findings are in contrast to  the DFT minimum energy structure reported in Ref.~\cite{Bodo12_13878}. Although in that case also an asymmetric HB network has been observed, where only two of the possible three HBs were formed, but there were no twofold HBs. This difference could be due to the finite temperature used in the present simulation.

\begin{figure}
\begin{center}
\includegraphics[width= \columnwidth]{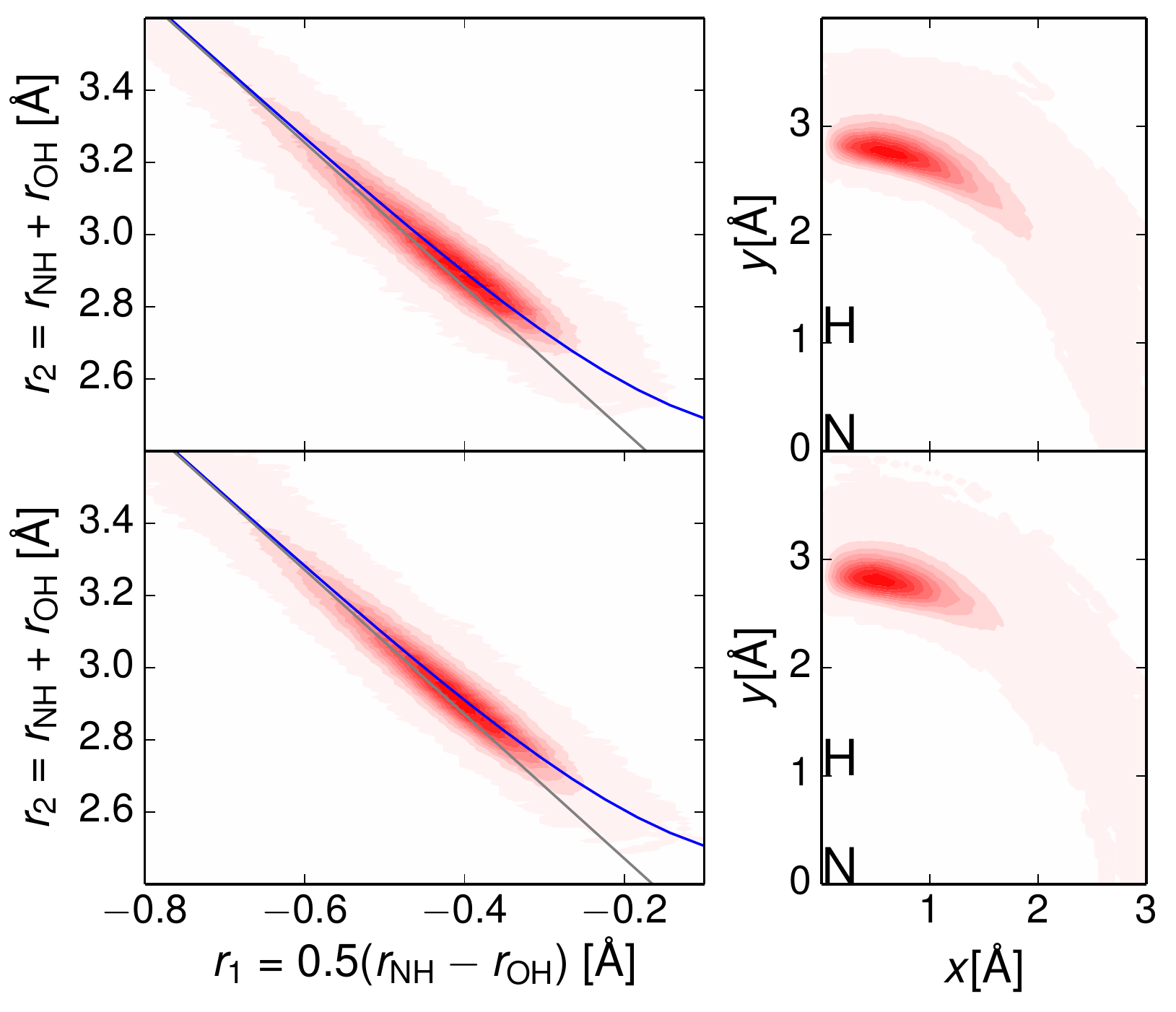}
\caption{Left: HB correlation plots from DFTB (top) and BLYP (bottom) trajectories averaged over all 6 pairs of along 60~ps NVT trajectory. Right: The position of the oxygen atom  involved in the HBs in the plane defined by the positions of the N, H and O atom. The N-H bond defines the $y$-axis and the origin is set to be at the nitrogen position.}
\label{fig:HBcorr}
\end{center}
\end{figure}

In the following the N-H$\ldots$O HBs are discussed, putting emphasis on geometrical correlations. To that end, HB coordinates $r_1 = 0.5(r_{\mathrm{NH}} -r_{\mathrm{OH}} ) $ and \mbox{ $r_2 = r_{\mathrm{NH}}  + r_{\mathrm{OH}} $} are introduced and plotted against each other in Fig.~\ref{fig:HBcorr} (left panel) for both methods. If the distances  $r_{\mathrm{NH}} $ and $r_{\mathrm{OH}}$ were fully uncorrelated, the results would lie on a straight line with the slope -2, plotted for reference in grey color. Within a HB  the H atom is pulled  towards the acceptor atom. Thus, $r_{\mathrm{OH}} $ is reduced as $r_{\mathrm{NH}} $ increases. This results in deviations from the linear relation between $r_1$ and $r_2$ and shifts the curve at constant $r_1$ to larger $r_2$ values. Furthermore, the closer $r_1$ is to zero, the stronger is the bond, as the hydrogen is located exactly between the two heavy atoms, corresponds to the strongest HB possible~\cite{limbach06_193}.

To compare observed correlations to other H-bonding systems, the valence bond model of Pauling is used. Here, starting from the HB distances, $r_{\mathrm{NH}}$ and $r_{\mathrm{OH}}$, bond orders are defined as $p_i = \exp(-(r_i - r_{i}^{\mathrm{eq}} )/b_i)$ with $i = \mathrm{ \{ NH, OH\} } $ \cite{pauling47_542}. Under the constraint that the sum of the two bond orders $p_i$ must be equal to one, the two coordinates depend on each other and the proton/H atom  transfer can be described by a single coordinate \cite{Limbach04_5195}. This path is drawn in Fig.~\ref{fig:HBcorr} as a blue line. The equilibrium distances, $r_i^{\mathrm{eq}}$, are obtained from single molecule gas phase geometry optimizations. The value for the  N-H bond is taken as the average of the three ammonia N-H bond lengths of the optimized structure, which are $ r_{\mathrm{NH}}^{\mathrm{eq},\mathrm{DFTB}}  = 1.029~\mathrm{\AA}$ and $r_{\mathrm{NH}}^{\mathrm{eq},\mathrm{BLYP}}= 1.036~\mathrm{\AA}$. The O-H bond length is $r_{\mathrm{OH}}^{\mathrm{BLYP}} = 0.991~\mathrm{\AA}$ in BLYP and $r_{\mathrm{OH}}^{\mathrm{DFTB}} = 0.982~\mathrm{\AA}$ in DFTB. The bond order decay parameters were changed simultaneously starting from the values reported for HBs in crystal structures \cite{steiner98_7041} until a good visual fit with the DFTB data was observed, resulting in $b_{\mathrm{OH}} = 0.3~\mathrm{\AA}$ and $b_{\mathrm{NH}} = 0.33~\mathrm{\AA}$. These values were not optimized separately for BLYP, as the DFTB results describe the BLYP correlations equally well. Overall, the HB distance correlations shown in Fig.~\ref{fig:HBcorr} (left panel) are strikingly similar in the two methods. In passing we note that a similar analysis was performed for trietylammonium nitrate using DFTB bulk MD simulations, where the resulting bond order decay parameters were found to be $b_{\mathrm{NH}} =0.355~\mathrm{\AA}$ and $b_{\mathrm{OH}}=0.321~\mathrm{\AA}$ \cite{zentel16_234504}.  Interestingly, judging from the geometric correlations, the HBs in the EAN cluster  are weaker than in the above bulk structure. 

Finally, we comment on the linearity of the HBs in the EAN cluster. Fig.~\ref{fig:HBcorr} (right panel) shows the distribution of O atom positions in the plane defined by the N, H, and O atom making the HB. Although being rather similar the deviations from linearity are more pronounced for DFTB as compared with BLYP. However, on average DFTB reproduces the EAN structures from  BLYP simulations very well as can be seen from the summary provided in Tab.~\ref{tab:rdfpeaks}.
\begin{figure}
\includegraphics[width=0.5 \textwidth]{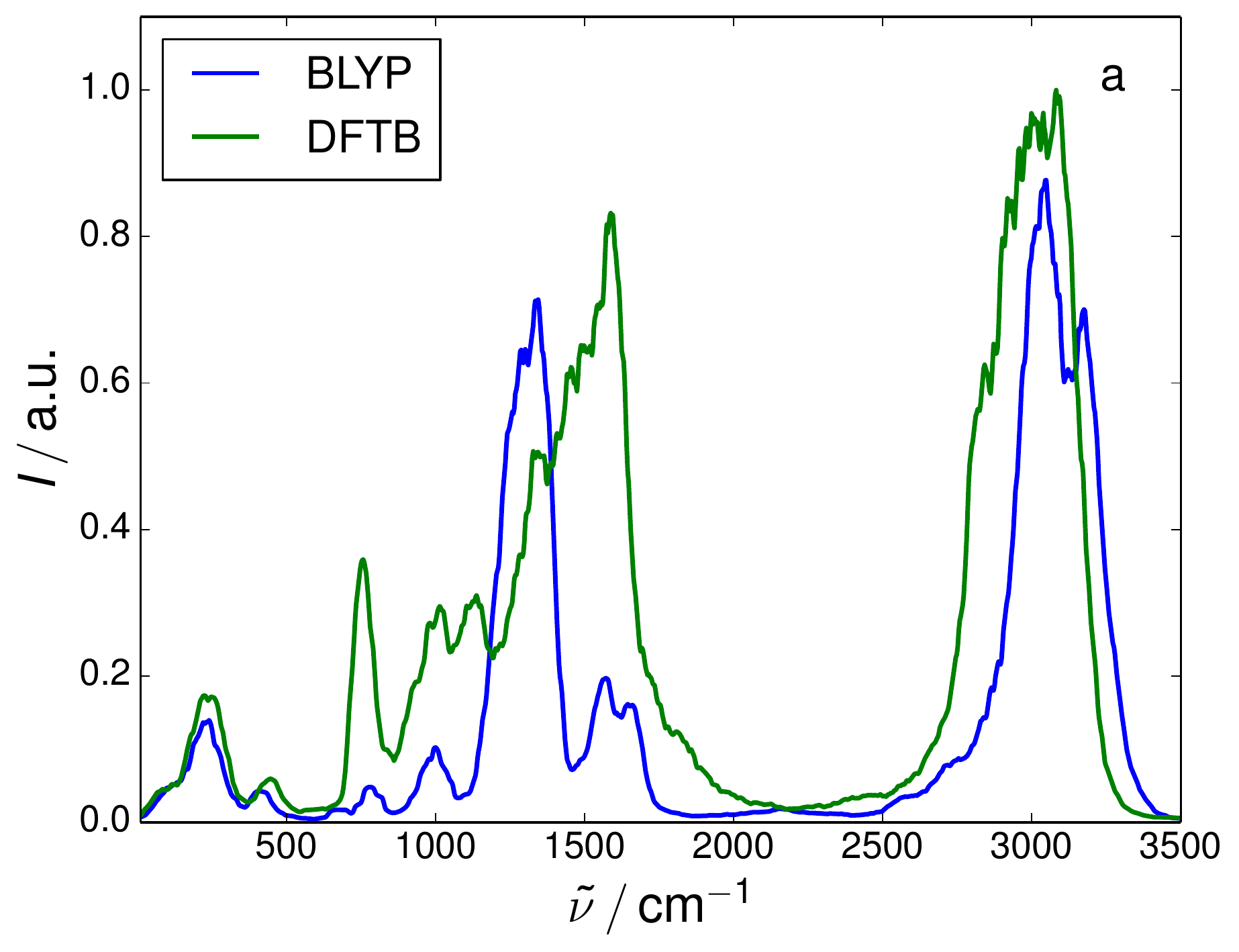} \\
\includegraphics[width=0.5 \textwidth]{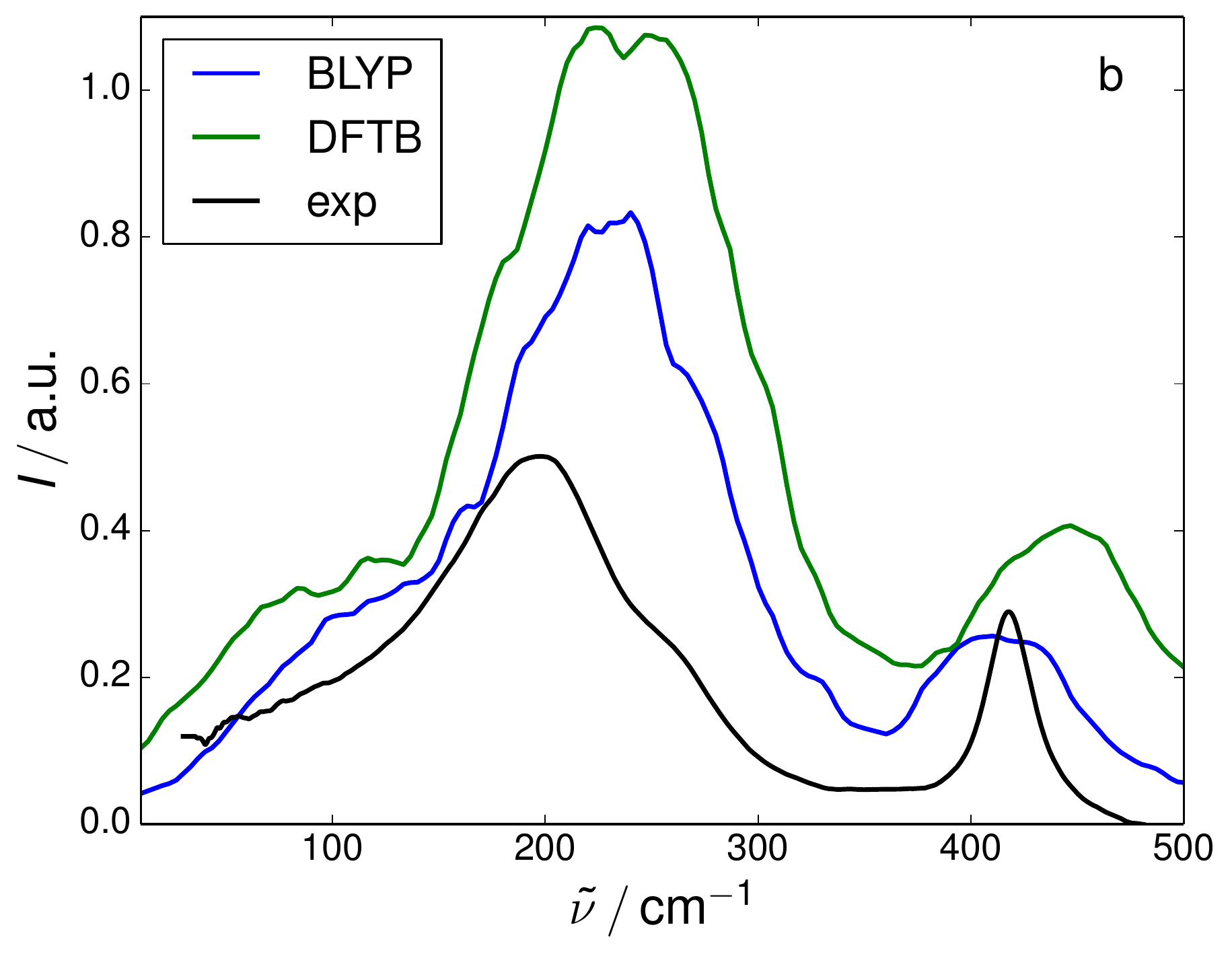} \\
\includegraphics[width=0.5 \textwidth]{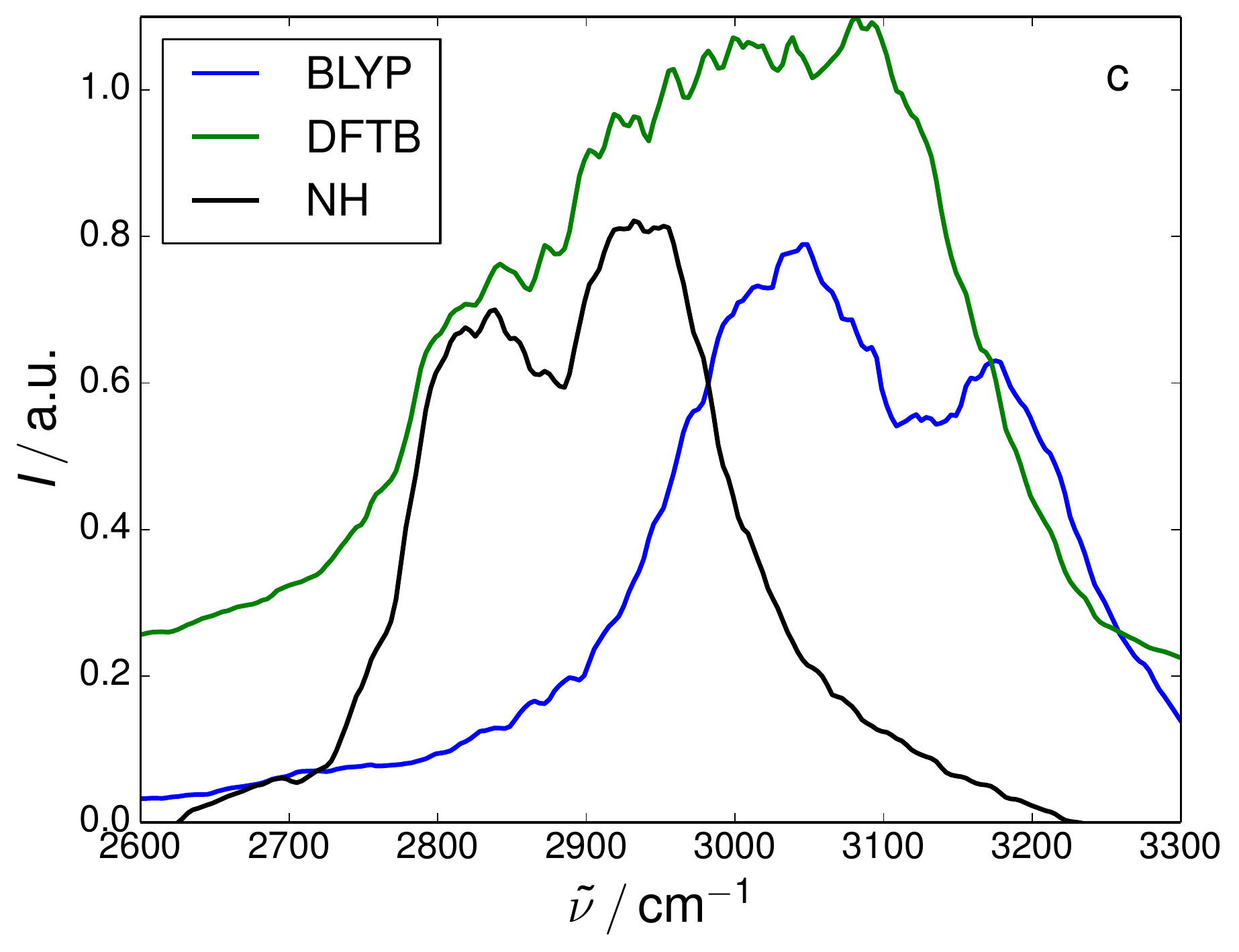} 
\caption{IR spectrum from the DFTB and BLYP trajectories of the cluster dipole from DFTB and BLYP dynamics in a) the whole IR range, b) a zoom into the FIR region (with offsets for better visibility) and c) the MIR  region, where also the contribution to the DFTB spectrum from the NH bond vibrations is given.
\label{fig:IRcluster}
}
\end{figure}
\subsection{IR Spectra} 
\begin{figure}
\begin{center}
\includegraphics[width= \columnwidth]{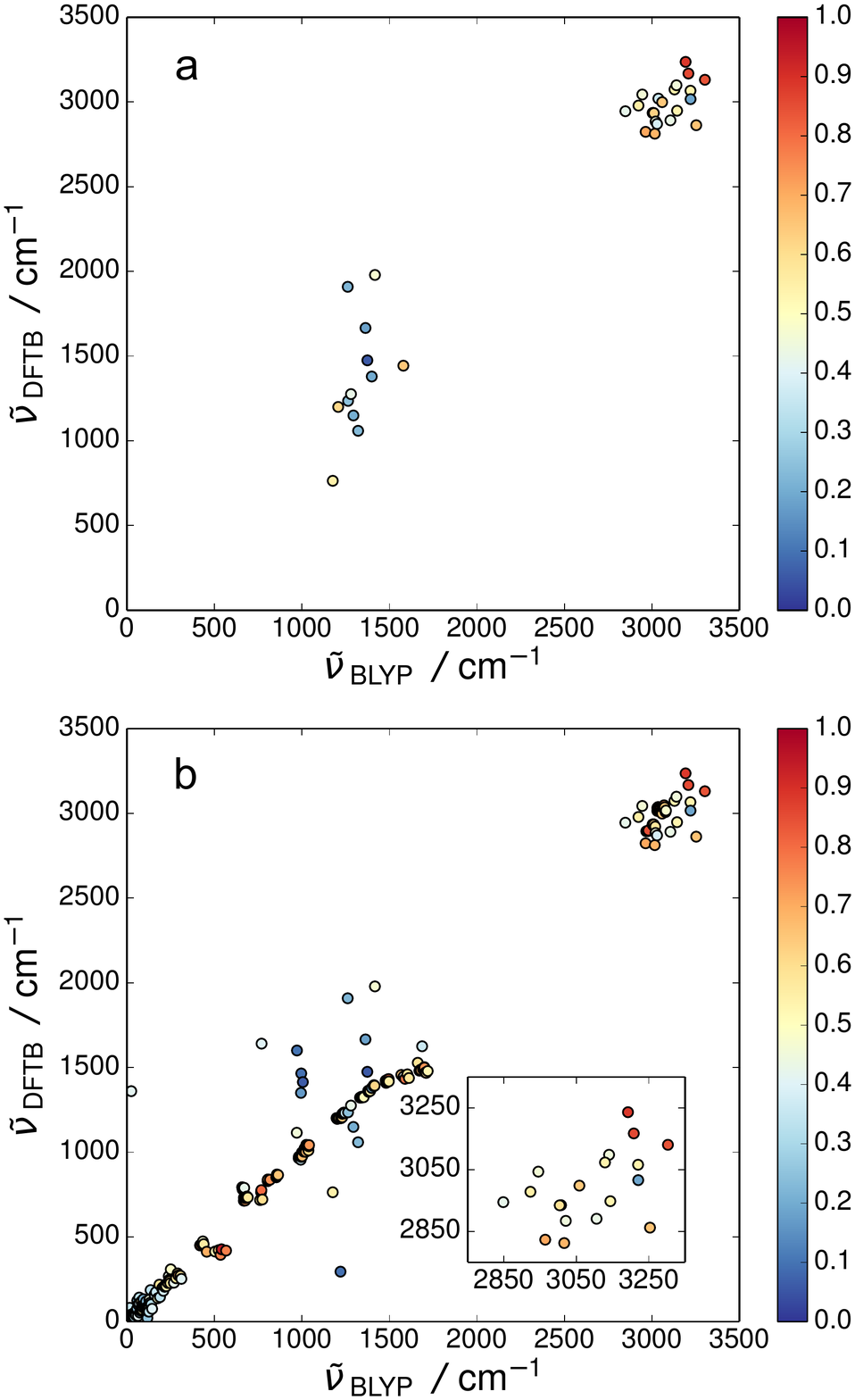}
\caption{(a) Overlap between DFTB and BLYP normal modes. The overlap between particular modes has been calculated such that the total overlap for all modes is maximized (Kuhn-Munkres algorithm~\cite{munkres57_32}). The outlier at about (20,1350)~\cm{} is considered an artefact of the assignment procedure. (b) Only those modes are shown where the associated BLYP IR transition has an intensity larger than 100 km/mole.}
\label{fig:Stickspectracluster}
\end{center}
\end{figure}

The IR spectra calculated  from BLYP and DFTB trajectories using Eq.~\eqref{eq:IRsignal} are shown in Fig.~\ref{fig:IRcluster}.  In panel (a) the complete spectral range is given, which reveals a qualitative agreement. That is,  most peaks are present in both spectra, but the intensities vary strongly. This discrepancy is particularly striking in the 750~$\mathrm{cm^{-1}}$ to 1900~$\mathrm{cm^{-1}}$ region.
To what extend the spectra deviate from one another due to differences in intensity only, or due to a different underlying nuclear dynamics, has been scrutinized using a  normal mode analysis. In  Fig.~\ref{fig:Stickspectracluster} DFTB and BLYP mode frequencies are correlated based on the overlap between the respective normal mode displacement vectors. Leaving aside the overlap, one finds a reasonable linear correlation for the range up to 900~\cm{} and for the N-H and C-H stretching range, i.e. here the density of states is rather similar for the two methods. Strong deviations are observed for the fingerprint region. Taking into account the overlap we notice the following: In the region up to about 350~\cm{} the overlap is in general below 50~\%, from 400 to about 900~\cm{} the overlap is considerably larger, in some case reaching almost 100~\%. In the region of the stretching vibrations (see inset of Fig.~\ref{fig:Stickspectracluster}), there is a reasonable overlap, but only the three NH-stretching modes with highest frequency reach about 100~\%. 
The problematic fingerprint region shows a good overlap for those modes, which match in frequency. In contrast, those modes which differ in frequency also have a small overlap. In the following we will compare the spectrum in the different spectral regions separately.

The FIR region shown in panel (b) is shaped by intramolecular vibrations as well as intermolecular dispersion and H-bonding  interactions. The experimental spectrum for bulk EAN from Ref.~\cite{fumino09_3184} is also given. It exhibits a double peak structure with a broad maximum at around 200~$\mathrm{cm^{-1}}$ and a smaller peak around 420~$\mathrm{cm^{-1}}$. Using DFT/B3LYP cluster calculations, Fumino \textit{et al.} \cite{fumino09_3184} assigned the low-frequency band to asymmetric and symmetric HB stretching vibrations at 197 and 128~$\mathrm{cm^{-1}}$, respectively, and the high-frequency band to a N-C-C deformation vibration of the cation. The shoulder around 250~$\mathrm{cm^{-1}}$ was found to be due to ethyl-nitrogen torsional motions. Note that this assignment was independent on the cluster size (from one to six pairs) and has been confirmed by the present BLYP normal mode calculations. In view of the correlation in Fig.~\ref{fig:Stickspectracluster} it is interesting to note that the spectrum as such is rather similar even though the overlap between the normal mode vectors is only modest.

Comparing theory and experiment, both DFTB and BLYP reproduce the observed double peak structure. The main peak is blue-shifted by about 25~$\mathrm{cm^{-1}}$ and its shoulder is not clearly resolved.  The small differences between theory and experiment could be due to the method  or due to different H-bonding in the cluster as compared with the bulk liquid. The high-frequency peaks are broader than in the  experiment, with DFTB predicting again a blue shift of about 25~$\mathrm{cm^{-1}}$. As far as the different widths are concerned one should note that in the gas phase cluster, the ethyl chains are oriented to the outer part of the structure and are therefore able to explore the phase space more freely than in a bulk system, which could lead to a broadening of the peak. 

H-bonding also manifests itself in red-shifted NH-stretching vibrational frequencies of those modes involved in a  HB with respect the free NH case. Details of the corresponding spectral region are shown in Fig.~\ref{fig:IRcluster}c. The IR spectrum obtained  from the DFTB trajectory shows a broad  plateau-like feature from 2900 to 3100~$\mathrm{cm^{-1}}$, while the DFT signal shows a double peak structure with a maximum at 3042 $\mathrm{cm^{-1}}$ and a distinct side band at 3166 $\mathrm{cm^{-1}}$. From the Mulliken charges of the DFTB simulation it is possible to extract just the dipole from a specific subgroup of atoms to better assign spectral features. The resulting signal  from the NH$_3$ group dipole is included in Fig.~\ref{fig:IRcluster}. The signal has two peaks at 2837 $\mathrm{cm^{-1}}$, and  2940 $\mathrm{cm^{-1}}$, and is very broad, an indication of HBs with various strengths. The remaining intensity in this range should be due to CH-stretching vibrations.

In order to judge the predictions in this spectral range, we compare with the Raman spectra of liquid EAN published in Ref.~\cite{Bodo12_13878} and analyzed based on hexamer DFT/$\omega$B97X-D normal mode calculations. In liquid EAN NH-stretching vibrations are found across the whole absorption band, \textit{i.e.} from 2900-3300~$\mathrm{cm^{-1}}$, reflecting the distribution of HB strengths. In addition, the cluster normal modes contain signatures of very strong HBs (around 2800~$\mathrm{cm^{-1}}$) and free NH-stretching vibrations (around 3400~$\mathrm{cm^{-1}}$). CH- stretching vibrations are assigned to the range between 2900-3000~$\mathrm{cm^{-1}}$). The observation that the low-energy part of the spectrum is dominated by H-bonded NH-stretching vibrations agrees with the present DFTB findings. Clearly, the finite size cluster cannot capture the broad distribution of HB strengths present in the liquid. In the BLYP case the absorption band is narrower hinting at less structural flexibility of the cluster (see also angular distribution in Fig.~\ref{fig:HBcorr}). Further, the first moment of the BLYP band appears to be blue-shifted as compared to DFTB. Given the fact that the highest frequency NH-stretching modes in the correlation plot, Fig.~\ref{fig:Stickspectracluster}, coincide, this evidences a more pronounced anharmonicity in DFTB as compared with BLYP.

Most interesting, however, is the range between 750~$\mathrm{cm^{-1}}$ and 1900~$\mathrm{cm^{-1}}$, which shows rather pronounced differences in Fig.~\ref{fig:IRcluster}(a). According to the Raman analysis in Ref.~\cite{Bodo12_13878} in this range one expects the NO$_3^-$ bending ($\sim$720~$\mathrm{cm^{-1}}$), symmetric stretching ($\sim$1045~$\mathrm{cm^{-1}}$), and asymmetric stretching ($\sim$1400~$\mathrm{cm^{-1}}$) vibrations. First, we assign this region of the IR spectrum according to the BLYP normal mode analysis as follows (in parentheses the position and intensity of the strongest peak) : 685 - 720~\cm{} -- NO$_3^-$ bending (717~\cm, 14 km/mole), 780 - 795~\cm{} -- NO$_3^-$ out-of-plane bending (780~\cm, 32 km/mole), 998 - 1032~\cm{} -- NO$_3^-$ symmetric stretching (998~\cm, 32 km/mole), 1489~\cm{} --  NO$_3^-$ asymmetric stretching (1489~\cm, 413 km/mole), 1561 - 1680~\cm{} -- NH$_3$ bending (1654~\cm, 152 km/mole). The range from 1002 - 1472~\cm{} contains rather mixed type of vibrations, involving mostly CH$_3$ and CH$_2$ bending motions and NO$_3^-$ asymmetric stretching vibrations. The largest intensity is at 1432~\cm (463 km/mole). Taking into account that the AIMD results contain the effects of anharmonicity and finite temperature, one could tentatively assign the spectrum in  Fig.~\ref{fig:IRcluster}(a)  as follows: 785~\cm{} -- NO$_3^-$ out-of-plane bending, 1000~\cm{} -- NO$_3^-$ symmetric stretching, double peak around 1320~\cm -- CH$_3$ and CH$_2$ bendings and NO$_3^-$ asymmetric stretching vibrations, double peak around 1580~\cm{} --  NH$_3$ bending vibrations. Compared to the Raman data there are some deviations, which are difficult to discuss since the experiment has been performed in the liquid phase. Decomposing the power spectrum of a Car-Parrinello simulation on the same size cluster Bodo \emph{et al.}~\cite{Bodo13_144309} obtained the NO$_3^-$ out-of-plane bending at 730~\cm, the NO$_3^-$ symmetric stretching at 962~\cm, and the NO$_3^-$ asymmetric stretching yielded a broad feature in the range 1280 - 1370~\cm. Overall, one can conclude that the present BLYP simulations are consistent with available experimental and simulation data. 

Apparently, the DFTB spectrum looks different in this spectral region and the question arises whether this is an effect of intensity, vibrational frequency or both. In order to scrutinize this issue, we have modified the correlation plot between DFTB and BLYP frequencies, taking into account only points where the BLYP intensity of the associated transition is above a certain threshold. If the threshold is such that all peaks in the fingerprint region are accounted for, the figure doesn't look much different from Fig.~
\ref{fig:Stickspectracluster}(a). The result shown in panel (b) of Fig.~
\ref{fig:Stickspectracluster} has been obtained for a threshold of 100 km/mole, which sets the focus on the strongest peak in the fingerprint region. From these observations we conclude that for the lower intensity peaks the vibrational frequencies match reasonably between BLYP and DFTB, but the intensities are different. For the region of the most intense band, which is due to CH$_3$ and CH$_2$ bending motions and NO$_3^-$ asymmetric stretching vibrations, both peak positions and intensities are different. In view of the favorable comparison of BLYP with previous simulation and experimental results, it must be concluded that DFTB does not yield a correct IR spectrum for the nitrate anion's vibrations. In passing we note that this finding is in accord with the conclusions drawn in Ref.~\cite{otte07_5751}, which reported deviation  between DFT and DFTB up to 10\% for NO-related vibrations of nitric acid.

\subsection{Charge Distributions}
The IR spectral regions of HB-related frequencies are well reproduced using DFTB. This suggestes to use the DFTB-Mulliken charges to  characterize HB-related atomic charge fluctuations in the thermal ensemble. First, we inspect the Mulliken charges of those atoms involved in H-bonding  given in Tab.~\ref{tab:averq}. They have been averaged over the whole trajectory and all H-bonded pairs. The DFTB values appear to be  closer to the force field parameters from Ref.~\cite{Umebayashi08_125} than to the BLYP ones. Closer inspection reveals that this is due to the combination of a small basis set and using Mulliken charges. Because a sufficiently large basis set is computationally not affordable for the BLYP simulations, we will consider only DFTB trajectory data in the following.

\begin{table}
\centering
\begin{tabular}{|c|c|c|c|}
\hline 
&  $\mathrm{N_{cation}}$ &\hspace{0.3cm} H\hspace{0.3cm} &\hspace{0.3cm} O\hspace{0.3cm}   \\ \hline 
DFTB  &  -0.14e & 0.28e & -0.73e   \\
\hline
BLYP & -1.15e & 0.53e & -0.01e  \\
\hline
FF \hspace{0.3cm}& -0.36e  & 0.31e  & -0.635e\\
\hline
\end{tabular}
\caption{Average  Mulliken charges along the DFTB and BLYP trajectories and partial charges of the force field (FF)  from Ref.~\cite{Umebayashi08_125}}
\label{tab:averq}
\end{table}

\begin{figure}
\begin{center}
\includegraphics[width=0.49 \textwidth]{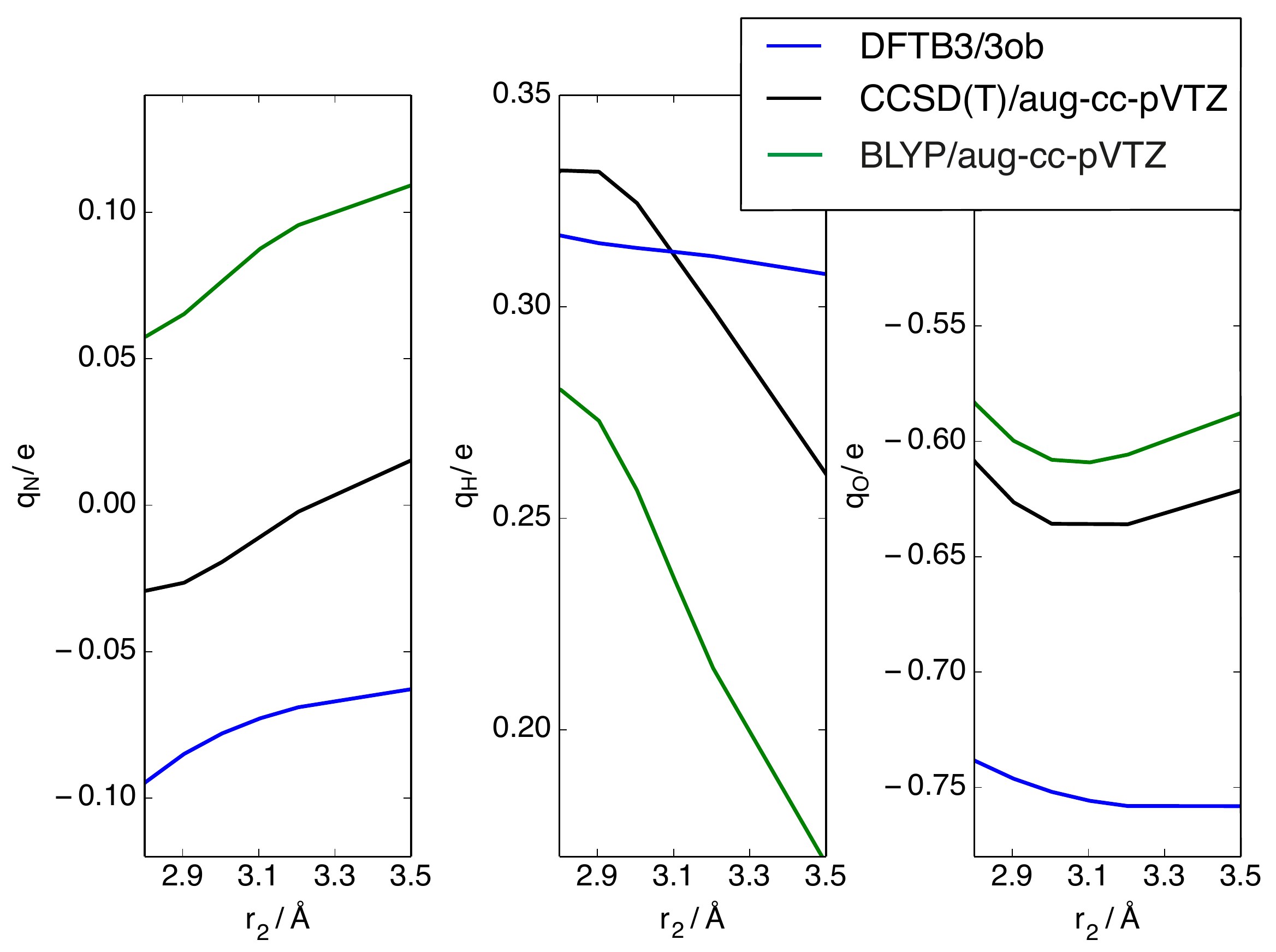} 
\caption{Dependence of Mulliken charges on the HB length, $r_2$, for a gas phase ion pair using different methods. The N-H bond length was kept constant at 1.04~\AA.}
\label{fig:charges_gas}
\end{center}
\end{figure}

In particular we will be interested in the distributions of  Mulliken charges on the N, O, and H atom of all HBs in the cluster with respect to the internal HB coordinates $r_1$ and $r_2$. In order to estimate the accuracy of DFTB in this respect, gas phase calculations of Mulliken charges for a single ion pair have been performed using different methods, i.e. DFTB, BLYP, and CCSD(T) (for basis sets, see figure). The resulting charges upon HB elongation are shown in Fig.~\ref{fig:charges_gas}. Taking CCSD(T) as a reference we notice that DFTB reasonably reproduces the CCSD(T) behaviour in case of the N atom, i.e. sign and slope match. For the O atom the sign matches as does the slope for short HB distances. The strongest deviation in slope is seen for the H atom. As far as BLYP is concerned, it is interesting to note that it fails to give the correct sign of the charge at the N atom. 

\begin{figure}
\begin{center}
\includegraphics[width=0.49 \textwidth]{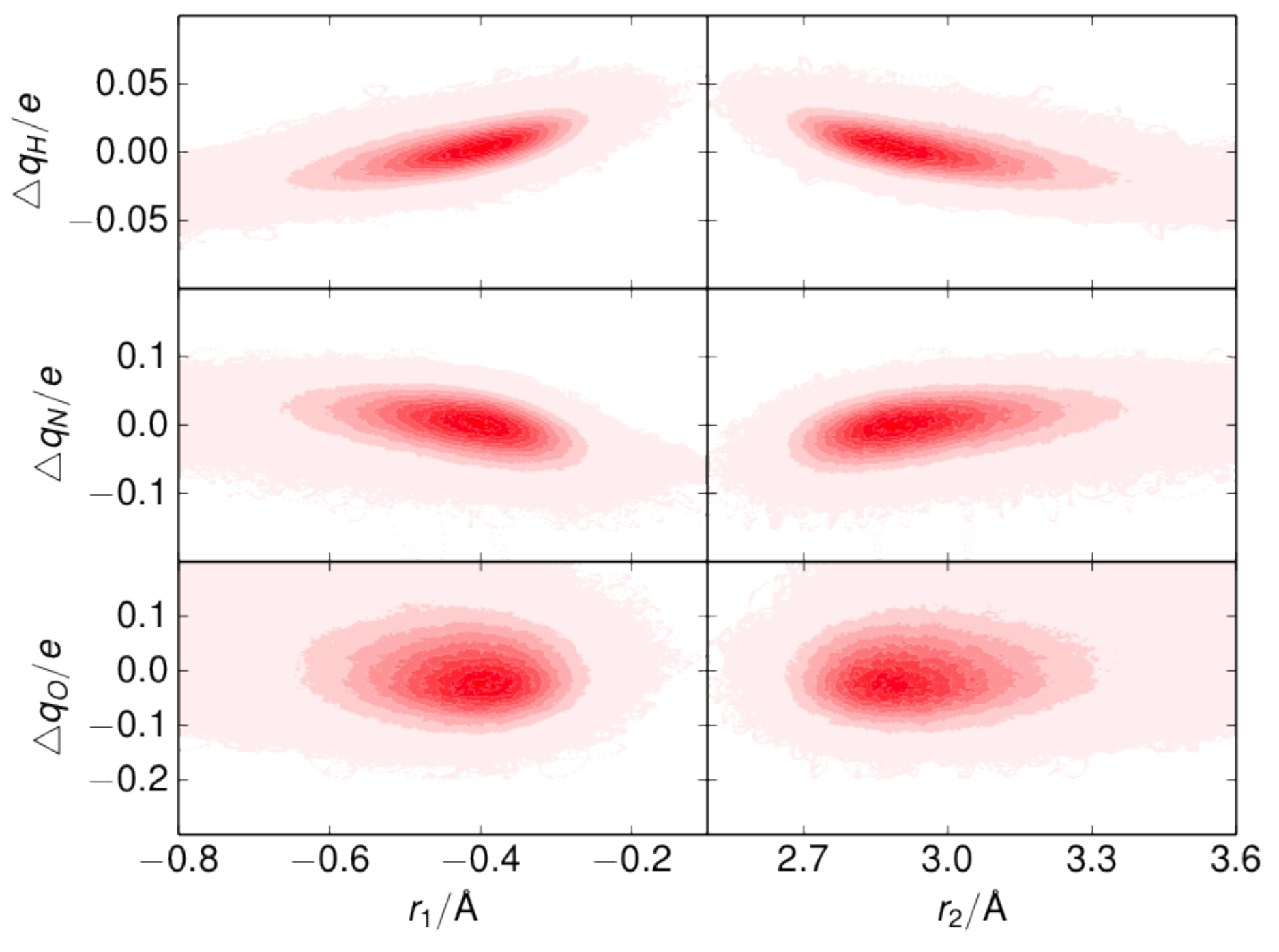} 
\caption{Density of the Mulliken charges (changes with respect to the average values)  along the HB coordinates for the DFTB trajectory. }
\label{fig:charges}
\end{center}
\end{figure}

In Fig.~\ref{fig:charges} we present the changes of Mulliken charges with respect to the average values sampled from the DFTB trajectory.
The following trends can be observed: The charge at the O atom shows a broad distribution with respect to both coordinates, but   no obvious correlation.  In contrast, for  the H and N atom the distributions are narrower and at the same time a coordinate dependence is clearly visible. The H atom's charge increases (i.e.\ less electron density) towards smaller absolut $r_1$ and smaller $r_2$, values, and at the same time the charge of the N atom decreases. As far as the H atom is concerned it can be anticipated, based on Fig.~\ref{fig:charges_gas} that the actual change should be even more pronounced.

\section{Conclusion} 
\label{sec:conclusions}
Molecular  Dynamics  simulations of a hexamer cluster of the protic ionic liquid EAN were performed with the goal being a critical assessment of the validity of the DFTB approach. To this end a comparison with DFT/BLYP as well as CCSD(T) results has been made. Averaged HB geometries agreed quantitatively for DFTB and BLYP  and  are in line with earlier experimental and theoretical publications.  With respect to the linearity of the HB BLYP predicts a slightly less flexible structure. HB coordinate correlations are rather similar for DFTB and BLYP, both being in the range of weak correlations. The IR spectra calculated from the cluster dipole trajectories show considerable differences between the two methods. In the far IR region the spectral features due to H-bonding are similar and close to available bulk experimental data. In the stretching region, DFTB yields a somewhat broader spectrum pointing to a larger flexibility of the cluster. Most critical is the fingerprint region, where DFTB clearly fails to give reasonable frequencies and intensities for the NO$_3^-$-related vibrations. A comparison of the normal modes between the two methods confirms this conclusion.

To sum up, DFTB can be considered as computationally less demanding alternative as compared to DFT if the focus is set on the properties and dynamics of HBs.

\begin{acknowledgements}
The authors thank the Deutsche Forschungsgemeinschaft (DFG) for financial support through the SFB 652 and A. Wulf for supplying the experimental FIR spectra.
\end{acknowledgements}


\begin{thebibliography}{10}
\providecommand{\url}[1]{{#1}}
\providecommand{\urlprefix}{URL }
\expandafter\ifx\csname urlstyle\endcsname\relax
  \providecommand{\doi}[1]{DOI \discretionary{}{}{}#1}\else
  \providecommand{\doi}{DOI \discretionary{}{}{}\begingroup
  \urlstyle{rm}\Url}\fi

\bibitem{Firaha15_7805}
D.S. Firaha, O.~Holl{\'o}czki, B.~Kirchner, Angew. Chem. Int. Ed. \textbf{54},
  7805 (2015)

\bibitem{Kerle09_12727}
D.~Kerl{\'e}, R.~Ludwig, A.~Geiger, D.~Paschek, J. Phys. Chem. B \textbf{113},
  12727 (2009)

\bibitem{Welton99_2071}
T.~Welton, Chem. Rev. \textbf{99}, 2071 (1999)

\bibitem{Zhang15_700}
Y.~Zhang, E.J. Maginn, J. Phys. Chem. Lett. \textbf{6}, 700 (2015)

\bibitem{hunt15_1257}
P.A. Hunt, C.R. Ashworth, R.P. Matthews, Chem. Soc. Rev. \textbf{44}, 1257
  (2015)

\bibitem{Fumino08_8731}
K.~Fumino, A.~Wulf, R.~Ludwig, Angew. Chem. Int. Ed. \textbf{47}, 8731 (2008)

\bibitem{Fumino12_6236}
K.~Fumino, E.~Reichert, K.~Wittler, R.~Hempelmann, R.~Ludwig, Angew. Chem. Int.
  Ed. \textbf{51}, 6236 (2012)

\bibitem{Roth12_105026}
C.~Roth, S.~Chatzipapadopoulos, D.~Kerl{\'e}, F.~Friedriszik, M.~L{\"u}tgens,
  S.~Lochbrunner, O.~K{\"u}hn, R.~Ludwig, New J. Phys. \textbf{14}, 105026
  (2012)

\bibitem{nibbering07_619}
E.T. Nibbering, J.~Dreyer, O.~K{\"u}hn, J.~Bredenbeck, P.~Hamm, T.~Elsaesser,
  in \emph{Analysis and {{Control}} of {{Ultrafast Photoinduced Reactions}}},
  ed. by O.~K{\"u}hn, L.~W{\"o}ste ({Springer Verlag}, Heidelberg, 2007), p.
  619

\bibitem{Zhuang09_3750}
W.~Zhuang, T.~Hayashi, S.~Mukamel, Angew. Chem. Int. Ed. \textbf{48}, 3750
  (2009)

\bibitem{Chatzipapadopoulos15_2519}
S.~Chatzipapadopoulos, T.~Zentel, R.~Ludwig, M.~L{\"u}tgens, S.~Lochbrunner,
  O.~K{\"u}hn, ChemPhysChem \textbf{16}, 2519 (2015)

\bibitem{Johnson13_224305}
C.J. Johnson, J.A. Fournier, C.T. Wolke, M.A. Johnson, J. Chem. Phys.
  \textbf{139}, 224305 (2013)

\bibitem{Bodo13_144309}
E.~Bodo, A.~Sferrazza, R.~Caminiti, S.~Mangialardo, P.~Postorino, J. Chem.
  Phys. \textbf{139}, 144309 (2013)

\bibitem{giese06_211}
K.~Giese, M.~Petkovi{\'c}, H.~Naundorf, O.~K{\"u}hn, Phys. Rep. \textbf{430},
  211 (2006)

\bibitem{mathias11_2028}
G.~Mathias, M.D. Baer, J. Chem. Theory Comput. \textbf{7}, 2028 (2011)

\bibitem{Thomas14_024510}
M.~Thomas, M.~Brehm, O.~Holl{\'o}czki, Z.~Kelemen, L.~Nyul{\'a}szi,
  T.~Pasinszki, B.~Kirchner, J. Chem. Phys. \textbf{141}, 024510 (2014)

\bibitem{may11}
V.~May, O.~K{\"u}hn, \emph{Charge and Energy Transfer Dynamics in Molecular
  Systems, 3rd revised and enlarged edition} ({Wiley-VCH}, Weinheim, 2011)

\bibitem{ivanov13_10270}
S.D. Ivanov, A.~Witt, D.~Marx, Phys. Chem. Chem. Phys. \textbf{15}, 10270
  (2013)

\bibitem{Maginn09_373101}
E.J. Maginn, J. Phys. Condens. Matt. \textbf{21}, 373101 (2009)

\bibitem{DelPopolo04_1744}
M.G. Del~P{\'o}polo, G.A. Voth, J. Phys. Chem. B \textbf{108}, 1744 (2004)

\bibitem{Bernardes14_6885}
C.E.S. Bernardes, K.~Shimizu, A.I.M.C. Lobo~Ferreira, L.M.N.B.F. Santos, J.N.
  Canongia~Lopes, J. Phys. Chem. B \textbf{118}, 6885 (2014)

\bibitem{Shimizu14_567}
K.~Shimizu, C.E.S. Bernardes, J.N. Canongia~Lopes, J. Phys. Chem. B
  \textbf{118}, 567 (2014)

\bibitem{zentel16_234504}
T.~Zentel, O.~K{\"u}hn, J. Chem. Phys. \textbf{145}, 234504 (2016)

\bibitem{Marx10_}
D.~Marx, J.~Hutter, \emph{Ab initio molecular dynamics: basic theory and
  advanced methods} ({Cambridge Univ. Press}, 2010)

\bibitem{kirchner15_202}
B.~Kirchner, O.~Holl{\'o}czki, J.N. Canongia~Lopes, A.A.H. P{\'a}dua, WIREs
  Comput. Mol. Sci. \textbf{5}, 202 (2015)

\bibitem{Zahn10_124506}
S.~Zahn, J.~Thar, B.~Kirchner, J. Chem. Phys. \textbf{132}, 124506 (2010)

\bibitem{Schroder11_024502}
C.~Schr{\"o}der, J. Chem. Phys. \textbf{135}, 024502 (2011)

\bibitem{Wendler12_1570}
K.~Wendler, M.~Brehm, F.~Malberg, B.~Kirchner, L.~Delle~Site, J. Chem. Theory
  Comput. \textbf{8}, 1570 (2012)

\bibitem{Thomas15_3207}
M.~Thomas, M.~Brehm, B.~Kirchner, Phys. Chem. Chem. Phys. \textbf{17}, 3207
  (2015)

\bibitem{Elstner98_7260}
M.~Elstner, D.~Porezag, G.~Jungnickel, J.~Elsner, M.~Haugk, T.~Frauenheim,
  S.~Suhai, G.~Seifert, Phys. Rev. B \textbf{58}, 7260 (1998)

\bibitem{Yang07_10861}
{Yang, Y}, H.~Yu, D.~York, Q.~Cui, M.~Elstner, J. Phys. Chem. A \textbf{111},
  10861 (2007)

\bibitem{Elstner06_316}
M.~Elstner, Theor. Chem. Acc. \textbf{116}, 316 (2006)

\bibitem{Addicoat14_4633}
M.A. Addicoat, R.~Stefanovic, G.B. Webber, R.~Atkin, A.J. Page, J. Chem. Theory
  Comput. \textbf{10}, 4633 (2014)

\bibitem{Bodo12_13878}
E.~Bodo, S.~Mangialardo, F.~Ramondo, F.~Ceccacci, P.~Postorino, J. Phys. Chem.
  B \textbf{116}, 13878 (2012)

\bibitem{Song12_2801}
X.~Song, H.~Hamano, B.~Minofar, R.~Kanzaki, K.~Fujii, Y.~Kameda, S.~Kohara,
  M.~Watanabe, S.i. Ishiguro, Y.~Umebayashi, J. Phys. Chem. B \textbf{116},
  2801 (2012)

\bibitem{fumino09_3184}
K.~Fumino, A.~Wulf, R.~Ludwig, Angew. Chem. Int. Ed. \textbf{48}, 3184 (2009)

\bibitem{zentel16_}
T.~Zentel, O.~K{\"u}hn, J. Mol. Liquids \textbf{226}, 56 (2016)

\bibitem{Humphrey96_33}
W.~Humphrey, {Dalke, A}, {Schulten, K.}, J. Molec. Graphics \textbf{14}, 33
  (1996)

\bibitem{Grimme10_154104}
S.~Grimme, J.~Antony, S.~Ehrlich, H.~Krieg, J. Chem. Phys. \textbf{132}, 154104
  (2010)

\bibitem{Gottwald12_}
F.~Gottwald, Bachlor thesis, University of Rostock (2012)

\bibitem{Ufimtsev09_2616}
I.~Ufimtsev, T.~Martinez, J. Chem. Theory Comput. \textbf{5}, 2619 (2009)

\bibitem{Gaus11_931}
M.~Gaus, Q.~Cui, M.~Elstner, J. Chem. Theory Comput. \textbf{7}, 931 (2011)

\bibitem{Gaus13_338}
M.~Gaus, A.~Goez, M.~Elstner, J. Chem. Theory Comput. \textbf{9}, 338 (2013)

\bibitem{Aradi07_5678}
B.~Aradi, B.~Hourahine, T.~Frauenheim, J. Phys. Chem. A \textbf{111}, 5678
  (2007)

\bibitem{Brehm12_5030}
M.~Brehm, H.~Weber, A.S. Pensado, A.~Stark, B.~Kirchner, Phys. Chem. Chem.
  Phys. \textbf{14}, 5030 (2012)

\bibitem{Ahlrichs89_165}
R.~Ahlrichs, M.~B{\"a}r, M.~H{\"a}ser, H.~Horn, C.~K{\"o}lmel, Chem. Phys.
  Lett. \textbf{162}, 165 (1989)

\bibitem{turbomole6.5}
{TURBOMOLE V6.5, a development of University of Karlsruhe and Forschungszentrum
  Karlsruhe GmbH, 1989-2007, TURBOMOLE GmbH, since 2007; available from
  http://www.turbomole.com}

\bibitem{Deglmann04_103}
P.~Deglmann, K.~May, F.~Furche, R.~Ahlrichs, Chem. Phys. Lett. \textbf{384},
  103 (2004)

\bibitem{Haser89_104}
M.~H{\"a}ser, R.~Ahlrichs, J. Comput. Chem. \textbf{10}, 104 (1989)

\bibitem{Hatting00_5154}
C.~H{\"a}tting, F.~Weigend, J. Chem. Phys. \textbf{113}, 5154 (2000)

\bibitem{Hayes11_3237}
R.~Hayes, S.~Imberti, G.G. Warr, R.~Atkin, Phys. Chem. Chem. Phys. \textbf{13},
  3237 (2011)

\bibitem{limbach06_193}
H.H. Limbach, G.S. Denisov, N.S. Golubev, in \emph{Isotope {{Effects}} in
  {{Chemistry}} and {{Biology}}}, ed. by A.~Kohen, H.H. Limbach ({CRC Press},
  Boca Raton, 2006), p. 193

\bibitem{pauling47_542}
L.~Pauling, J. Am. Chem. Soc. \textbf{69}, 542 (1947)

\bibitem{Limbach04_5195}
H.H. Limbach, M.~Pietrzak, S.~Sharif, P.M. Tolstoy, I.G. Shenderovich, S.N.
  Smirnov, N.S. Golubev, G.S. Denisov, Chem. Eur. J. \textbf{10}, 5195 (2004)

\bibitem{steiner98_7041}
T.~Steiner, J. Phys. Chem. A \textbf{102}, 7041 (1998)

\bibitem{munkres57_32}
J.~Munkres, J. Soc. Ind. Appl. Math. \textbf{5}, 32 (1957)

\bibitem{otte07_5751}
N.~Otte, M.~Scholten, W.~Thiel, J. Phys. Chem. A \textbf{111}, 5751 (2007)

\bibitem{Umebayashi08_125}
Y.~Umebayashi, W.L. Chung, T.~Mitsugi, S.~Fukuda, M.~Takeuchi, K.~Fujii,
  T.~Takamuku, R.~Kanzaki, S.i. Ishiguro, J. Comput. Chem. Jpn. \textbf{7}, 125
  (2008)

\end{thebibliography}
\end{document}